\documentstyle[12pt,aaspp4]{article} 
\begin{document} 

\received{~~} \accepted{~~} 
\journalid{}{}
\articleid{}{}



\baselineskip=14pt 
\def\gsim{ \lower .75ex \hbox{$\sim$} \llap{\raise .27ex \hbox{$>$}} } 
\def\lsim{ \lower .75ex\hbox{$\sim$} \llap{\raise .27ex \hbox{$<$}} } 
\input{psfig.tex}

\title{Redshift determination in the X--ray band of gamma--ray bursts}

\author{Gabriele Ghisellini}
\affil{Osservatorio Astronomico di Brera, Via Bianchi 46, I--23807 Merate, Italy} 
\author{Francesco Haardt} 
\affil{Dipartimento
di Fisica dell'Universit\`a di Como, via Lucini 3, I--22100 Como, Italy }
\author{Sergio Campana, Davide Lazzati$^1$, Stefano Covino} 
\affil{Osservatorio Astronomico di Brera, Via Bianchi 46, I--23807 Merate, 
Italy \\ 
$^1$ Dipartimento di Fisica dell'Universit\`a di Milano, 
Via Celoria 16, I--20133 Milano, Italy}

\begin{abstract}

If gamma--ray bursts originate in dense stellar forming regions, the
interstellar material can imprint detectable absorption features on the observed
X--ray spectrum.  Such features can be detected by existing and planned X--ray
satellites, as long as the X--ray afterglow is observed after a 
few minutes from the burst.  
If the column density of the interstellar material exceeds $\sim
10^{23}$ cm$^{-2}$ there exists the possibility to detect the $K_\alpha$
fluorescent iron line, which should be visible for more than one year, long
after the X--ray afterglow continuum has faded away.  
Detection of these X--ray features will 
make possible the determination of the redshift of gamma--ray bursts
even when their optical afterglow is severely dimmed by extinction.

\end{abstract}

\section{Introduction}

After the breakthrough of BeppoSAX in gamma--ray burst science (Costa et al.
1997; Van Paradijs et al.  1997), to solve the mystery of the origin
of gamma--ray bursts we still
need to build up a sizeable sample of objects with known redshift.  Optical
spectroscopy is however difficult, as both the optical afterglow and the
presumed host galaxy are faint.  Moreover, the optical afterglow is not observed
in nearly half of the bursts with observed X--ray afterglows:  a possible
explanation is optical extinction, favoring models in which the bursts are
formed in dense regions of stellar formation (Paczy\'nski 1998).

If this is the case, the typical densities (of hydrogen atoms) is of the
order of $n\sim 10^4$--$10^5$ cm$^{-3}$, with typical sizes of tens of parsecs.
It is therefore conceivable to have column densities $N_{\rm H} \gsim 10^{22}$
cm$^{-2}$, which would, besides heavily reddening the optical emission of the
bursts, also absorb their X--ray spectra.  GRB970828 observed by ASCA (Murakami
1998) and GRB980329 observed by BeppoSAX (in't Zand et al.  1998) have already
given some evidence of intrinsic high column densities.
The possibility exists, therefore, to observe
absorption edges in the X--ray spectra of gamma--ray burst afterglows.
This would lead to the determination of the redshift using only X--ray
data, as long as the observations are carried out when the decaying X--ray
afterglow is sufficiently bright.  
Along these lines, 
Loeb \& Perna (1998) discussed the time behavior of optical absorption lines,
and M\'esz\'aros \& Rees (1998) discussed the possibility to observe broad
absorption features originating in the relativistic fireball, and
qualitatively investigated the detectability of 
iron emission lines in the X--ray spectra of bursts.

In this paper we investigate the detectability of X--ray features in detail,
finding which part of the $N_{\rm H}$--redshift plane can be accessed by present
and future X--ray missions.

If bursts (or a fraction of them) originate in dense regions, then also the
re-emission of the absorbed energy should be important.  In the X--ray band, the
most prominent emission feature will probably be the fluorescent $K_{\alpha}$
iron line.  We then calculate the expected flux of this line and its light
curve, to see if and when it can be detected.

\vskip 1 true cm

\section{The $N_{\rm H}$--redshift plane} \vskip 0.5 true cm In Fig.  1 we show
the X--ray spectra observed assuming different values of $N_{\rm H}$, located at
$z=0$, and an intrinsic power law photon spectrum $\propto \epsilon^{-1}$, where
$\epsilon$ is the photon energy.  The cut--off frequency produced by absorption
gives a relation between $N_{\rm H}$ and $z$, since a small cut--off frequency
can be produced by a small column at a small redshift, or a large column at a
large redshift.  To disentangle this degeneracy we must detect some absorption
edges.

The absorption edges are more prominent for larger $N_{\rm H}$, but the observed
flux is dimmer.  The detectability of a particular edge will therefore be
possible only in a range of $N_{\rm H}$, depending on the particular considered
edge, the intrinsic (unabsorbed) flux of the burst, the integration time and the
effective area of the detector.

Let $C_1^\prime$ and $C_2^\prime$ be the number of counts in the energy bins
just below and just above an absorption edge.  If $C_1^\prime$ and $C_2^\prime$
are independent measures of the number of success in two stochastic processes,
we can say that $C_1^\prime$ differs from $C_2^\prime$ at a level of $f$ sigma
if:

\begin{equation} 
C_1^\prime-C_2^\prime\, > \,f \,\sqrt{C_1^\prime+C_2^\prime}
\end{equation}

The observed counts are related to the `unabsorbed' counts (that we would have
observed with  $N_{\rm H}=0$) by 
$C_1^\prime = C_1 e^{-\tau_1}$ and 
$C_2^\prime =C_2 e^{-\tau_2}$, 
where $\tau_1$ and $\tau_2$ are the absorption optical depths
at energies just below and above the edge.

In addition, since the two energy bins are centered at the two energies
$\epsilon_1$ and $\epsilon_2$, we have $C_2=C_1
(\epsilon_2/\epsilon_1)^{-\Gamma}$, where $\Gamma$ is the photon spectral index.
Then we have

\begin{equation} 
C_1 \, > \, f^2 \, { e^{-\tau_1}+(\epsilon_1/\epsilon_2)^\Gamma e^{-\tau_2} 
\over \left[ e^{-\tau_1}- (\epsilon_1/\epsilon_2)^\Gamma 
e^{-\tau_2} \right]^2 } 
\end{equation}

The optical depth below and above the edge can be approximated by a polynomial
function (e.g.  Balucinska--Church \& McCammon 1992).  
As an example, the optical
depths below and above the neutral oxygen (at 0.521 keV) 
and neutral iron (at
7.111 keV) edges, for solar abundances, are 
$\tau_{1,O}\simeq 652.3 \, \tau_T$; 
$\tau_{2,O} \simeq 1095.2\, \tau_T$; 
$\tau_{1,Fe}\simeq 1.265\, \tau_T$; 
$\tau_{2,Fe}\simeq 2.360\, \tau_T$,
where $\tau_T$ is the Thomson scattering optical depth.

In Fig.  2 we show the minimum number of counts (i.e.  $C_1$) necessary to
detect the neutral oxygen and the neutral iron edges, as a function of $N_{\rm
H}$, calculated as above.  
The solid lines in Fig. 2 correspond to Equation (2)
(for $f=$1, 2, 3), while dashed lines correspond to the actual number of counts
detected, taking into account absorption (i.e.  they correspond to $C^\prime=C
e^{-\tau_1}$).  It can be noted that this simple estimate refers only to the
number of counts in two bins (immediately below and immediately above) the edge
energy.  
However, we should add also the information contained into the other
adjacent bins, increasing the detectability of the edge.  We can therefore
consider Equation (2) and Fig.  2 as a conservative, albeit rough, estimate.

Fig. 2 shows the range of $N_{\rm H}$ at which a particular detector is
sensitive.  For a given effective area, exposure time and $unabsorbed$ flux, we
can estimate the number of counts that would be detected (in the absence of
absorption).  This can be represented by an horizontal line on Fig.  2,
intercepting the solid lines in two points, which defines the extremes of the
range of $N_{\rm H}$ for which we can see the considered edge.  The actual
observed counts (including absorption) are then given by the dashed lines.

Analogously, Fig.  2 can be used to estimate the exposure time needed to detect
the edge once we know the observed (absorbed) flux (dashed lines).  This is
given by the dashed lines in Fig.  2, by dividing the known flux by the required
counts and by the effective area.  In this case, the solid lines give the
corresponding unabsorbed flux.

\section{Simulations}

Let $A$ be the effective area of the X--ray detector, and $F$ the integrated
flux of the burst in a given energy band, between $\epsilon_1$ and $\epsilon_2$.
In general, $A$ will be a function of frequency, while $F$ a function of time.
Although we still lack information on the detailed behavior of the X--ray flux
soon after the burst, let assume a power law behavior, connecting the flux at
the time of burst and the afterglow as seen a few hours later.  We know that
$\sim 10$ hours after the burst, the observed 2--10 keV flux is of the order of
$\sim 10^{-13}$ erg cm$^{-2}$ s$^{-1}$  (in't Zand et al. 1998;
Frontera et al. 1998; Piro et al. 1998a, 1998b).
Assuming a $t^{-1}$
dependence, we obtain a flux of $F(t=100~{\rm s})=3.6 \times 10^{-9}$ erg
cm$^{-2}$ s$^{-1}$ during, or immediately after, the burst [corresponding to a
monochromatic flux of $160\,\mu$Jy if the energy spectral index is flat
($\Gamma=1$)].

For $\Gamma=1$ the counts detected in $t_{exp}$
exposure time, starting at $t_1$, in the energy bin $\Delta \epsilon$ are

\begin{eqnarray} 
C^\prime \, &=&\, 
A \, F(t_1)
\,{\Delta \epsilon\over \epsilon}\, 
{1 \over \epsilon_2-\epsilon_1} \ln (t_{exp}/t_1) \nonumber \\ 
&\approx & \, 62.4 \, 
\left( {A\over 100~{\rm cm^2}}\right) 
\left( {F(t_1) \over 10^{-8}~{\rm erg~cm^{-2}~s^{-1}}}\right)
\left( {\Delta \epsilon/\epsilon \over 0.1}\right) 
{\ln (t_{exp}/t_1) \over \epsilon_{2,{\rm keV}}-\epsilon_{1,{\rm keV}}} 
\end{eqnarray}

Given the assumed time decay law and energy spectrum, the necessary S/N ratio
can be achieved only if the X--ray observation starts immediately after the
burst itself, or if the collecting effective area of the detector is much larger
than 100 cm$^2$.

We have simulated the observed spectrum by using the response matrices of some
future planned missions, such as JET-X ($A\sim 200$ cm$^2$ at 1.5 keV and 
$A\sim 40$ cm$^2$ at 8.1 keV for the two telescopes; Citterio et al. 1996), 
AXAF with
Back Illuminated (BI) CCDs ($A\sim 700$ cm$^2$ at 1.5 keV and $A\sim 40$ cm$^2$
at 8.1 keV; Kellogg et al.  1997) and XMM with the EPIC detectors ($A\sim 3600$
cm$^2$ at 1.5 keV and $A\sim 1500$ cm$^2$ at 8.1 keV for three telescopes;
Gondoin et al.  1996) and assuming:  
i) $F(t_1)=10^{-8}$ between 2 and 10 keV at
the beginning of the observation; 
ii) a power law time decay of the flux
$\propto t^{-1}$; iii) an intrinsic (unabsorbed) power law spectrum of photon
index $\Gamma=1$ in the considered energy range, constant in time.  We
considered these three telescopes because they cover a wide range of effective
areas, have a good spectral resolution and allow for a low background
contamination.

We simulated two different cases: a GRB afterglow at $z=0.25$ and 
intrinsic $N_{\rm H}=3\times 10^{21}$ cm$^{-2}$ 
and $z=4$ and $N_{\rm H}=10^{24}$ cm$^{-2}$, which
are relevant for the oxygen and iron edge, respectively.  
A galactic column
density of $3\times 10^{20}$ cm$^{-2}$ has also been included.

In the case of the oxygen edge (at 0.52 keV) the 
satellite energy band is extremely
important in order to recover the correct GRB redshift.  
In the case of JET-X,
the minimum energy of 0.3 keV limits the maximum detectable redshift to $\sim
0.7$.  The influence of the galactic absorption plays also a crucial role, such
that only for low values ($\lsim\, 5\times 10^{20}$ cm$^{-2}$) we are able to
disentangle the intrinsic and galactic absorption.  
In Fig. 3 we report the
contour plots in the $N_{\rm H}-z$ plane of the simulated models as observed
with different X--ray satellites.  The three contours refer to 1, 2 and
$3\,\sigma$ confidence levels.  
In Fig. 3a is shown the case of the JET-X telescope.  It
can be noted that the input redshift and column density are not recovered
satisfactorily.  In particular, the presence of different absorption features
(O, Ne, Mg, Si) results in the elongated contour in the $N_{\rm H}-z$ plane.  
In
the case of AXAF (Fig. 3b), the recovery of the GRB redshift is eased by the
higher throughput at low energies guaranteed by the BI CCDs.  
The large effective
area of XMM poses no problem for the identification of the redshift (Fig. 3c).

In the case of the Fe edge there are less problems due to the fact that beyond
iron there are not prominent K edges.  This is testified by Fig. 4, in which
for all the considered instrument the redshift and the column density are
recovered with a high degree of confidence.

\section{Re-emission of the fluorescent $K_\alpha$ iron line}

The dense material responsible for the absorption we have discussed in the
previous sections will re-emit the absorbed energy, and part of it will be
re-emitted in the form of X--ray lines.  Among those, the fluorescent $K_\alpha$
iron line will be the most prominent.  A very crude estimate of the expected
flux can be done by assuming that a fraction $a$ of the fluence ${\cal F}$
emitted by the burst is re-emitted by a region of dimension $R$, in an observed
time-scale of the order of $R/c$.  Then the average flux of the iron fluorescent
$K_\alpha$ line is

\begin{equation} 
F_{K_\alpha} \, \approx \, {a\, {\cal F} \over R/c} \, 
\approx\, 10^{-16}\, \left( {a \over 0.01}\right) \,\, 
\left({{\cal F}\over 10^{-5} {\rm
\,erg~cm^{-2}}}\right)\, \left( {1~{\rm pc}\over R}\right) 
\,\, {\rm erg~cm^{-2}~s^{-1}} 
\end{equation}

\noindent This promising result motivates a somewhat deeper analysis.

At the energy of the neutral iron edge and assuming a solar abundance of iron,
the optical depth for photoelectric absorption is $\tau_{7.1}\sim 2.36 \,\tau_T$
and increases somewhat for partially ionized iron.  The incoming ionizing flux
preferentially photoionizes an electron of the iron $K$ shell, instead of
electrons of outer shells, because lower energy photons are preferentially
absorbed by lighter elements.  The absorption is immediately followed by
fluorescence emission of a $K_\alpha$ photon whose energy ranges between 6.4 and
6.9 keV, depending on the ionization state.  An initially neutral iron atom can
therefore absorb up to a maximum of 26 ionizing photons.  Setting $n$ the number
density of hydrogen, the absorption optical depth for solar abundance, as long
as the iron atoms are not completely ionized, is

\begin{equation} 
\tau_{edge}\, \approx \, 0.5 \left( {n \over 10^5~{\rm cm^{-3}}} \right) 
\, \left( { R\over 1~{\rm pc} }\right) 
\end{equation}

If the bursts are located in a dense star forming region it is very likely that
a significant fraction of the energy above the iron edge is absorbed.

Assuming that a fraction $\tau_{edge}$ (with $\tau_{edge}\le 1$) of all photons
between $\epsilon_{edge}$ and $2\,\epsilon_{edge}$ get absorbed, the energy
re-emitted in the $K_{\alpha}$ line is a fraction
$Y\tau_{edge}\epsilon_{edge}/\epsilon_{max}$ of the total energy of the burst,
where $Y$ is the yield (equal to 0.37 for neutral iron, and larger for partially
ionized iron), $\epsilon_{edge}$ is the edge energy, and $\epsilon_{max}$ is the
energy where the burst spectrum peaks [in a $\epsilon F(\epsilon )$
representation].  The average flux received in the iron line is

\begin{eqnarray} 
F_{K_\alpha}\, & \simeq &\, Y \tau_{edge} \, {\epsilon_{edge}
\over \epsilon_{max}} { {\cal F} \over R/c} \nonumber \\ 
&\approx & \, 5.8\times
10^{-16} \left( {Y\over 0.5} \right) \left( {n\over 10^5{\rm\, cm^2} } \right)
\left( {{\cal F} \over 10^{-5} {\rm \,erg~cm^{-2}} }\right) \left(
{\epsilon_{edge}/\epsilon_{max} \over 0.025 }\right) 
\quad {\rm erg~cm^{-2}~s^{-1}} 
\end{eqnarray}

The line flux does not explicitly depend upon the dimension $R$, which however
controls the duration of the line emission ($\sim R/c$).  This dimension can be
estimated by equating the number of ionizing photons to the total number of iron
atoms present in a sphere of radius $R$, assuming that each iron atom absorbs 26
ionizing photons and a solar abundance $x_{Fe}$ of iron:

\begin{equation} 
N_{edge}\, \simeq \, {E \over \epsilon_{max}}\, 
{\Delta \epsilon\over \epsilon_{edge}}\, \simeq \, 
26\, x_{Fe} \ n {4\pi R^3\over 3}
\end{equation} \begin{equation} R\, \approx \, 
0.65 \left[ \left({E_{52}\over \epsilon_{300}}\right)\, 
\left( {\Delta\epsilon\over \epsilon_{edge}}\right)\,
\left( {3\times 10^{-4} \over x_{Fe}}\right)\, 
\left( {10^5{\rm \, cm^{-3}}\over n}\right)\right]^{1/3} 
\quad {\rm pc} 
\end{equation} 
where $E=10^{52}\,E_{52}$
erg is the total burst energy, whose spectrum peaks at
$\epsilon_{max}=300\,\epsilon_{300}$ keV.  
Note that with $n=10^5\,$cm$^{-3}$
and $R$ given by Equation (12) one obtains a column of $N_{\rm H}\sim 2\times
10^{23}$ cm$^{-2}$.  
The $K_\alpha$ iron line calculated in this simplified way is 
approximately constant for 2 years (but see below), 
long after the X--ray afterglow has decayed below the
detection threshold (the 2--10 keV X--ray flux should be 10$^{-16}$ erg
cm$^{-2}$ s$^{-1}$ after a month, assuming a flux of $10^{-13}$ erg cm$^{-2}$ s
$^{-1}$ after $\sim 10$ hours).  The line flux should be narrowly concentrated
in the observed energy range $[6.9-6.4]/(1+z)$ keV.
In the following section we present a detailed calculation
for the light curve behavior of the line flux, including
self consistently the effect of the absorption of the ionizing flux.

\subsection{The light curve of the iron fluorescence line}

Emission of fluorescence line photons can be treated as a scattering process,
with each iron atom scattering off 26 ionizing photons.  In order to compute the
light curve of the scattered radiation in a spherical geometry, assuming that
the ionizing photons come from a point--like central source, one has to consider
the effects of the time--delays induced by geometry.  
A comprehensive treatment
of a similar problem has been done by Blandford \& McKee (1992),
in the context of reverberation of the broad line region of AGNs, 
responding to a variable photoionizing central source.
They however neglected the absorption term of the ionizing flux,
which instead is important here.
Assuming, for semplicity, an infinite medium, it is not difficult to show that 
the flux of scattered photons is given by

\begin{equation} 
\dot N_{\rm line}(t) \,= 
2\pi Y \int_{-1}^{+1} d\mu \,\,
\int_{R_{\rm min}}^{R_{\rm max}} dr \, r^2 n_F(r)\, 
\int_{\nu_0}^{\infty} {d\nu \over h\nu} \, 
{L_{\nu}(t') \over 4\pi r^2} \, \sigma_{\nu}\, {\rm
e}^{-\int_0^r dr^\prime n_F(r^\prime)\sigma_{\nu}} 
\end{equation}
where $r$ is the radial distance of the
scattering event from the central source and
$\mu$ is the cosine of the scattering angle.
The values of $r$ and $\mu$ are related to the arrival time $t$ 
through $t=t^\prime+(1-\mu)\, r/c$, where 
$t^\prime$ is the time of photons arriving directly, without
being scattered.
$L_{\nu}$ is the source specific luminosity.  
The factor $Y$ is the fluorescence yield, $n_F$ is
the scattering material number density, i.e., in our case, it is 26 times the
iron number density and $\sigma_\nu$ is the photoelectric cross section.
The minimum and maximum value of the integration radius are given by 
the conditions 
$t'\ge 0$ and 
$t'\le \Delta t$, 
where we assumed the source light curve to be steady for a duration $\Delta t$.  
Under such assumption we have 
$R_{\rm min}={\rm max}[0, c\, (t-\Delta t)/(1-\mu)]$ and 
$R_{\rm max}=c\, t/(1-\mu)$.

The above integral can be worked out if one neglects the frequency
dependence of the cross section and of the ionizing flux, and
assuming further an homogeneous scattering medium.
In this case
\begin{equation}
\int_{\nu_0}^{\infty} d\nu {L_{\nu}(t') \over h\nu} \, \sigma_{\nu}\, 
{\rm e}^{-\int_0^r dr^\prime n_F(r^\prime)\sigma_{\nu}} \, =\, 
{ \dot N_{\rm ion} \over h} { {\rm e}^{-r/r_F} \over r_F n_F}
\end{equation}
where 
the scale radius $r_F=1/(n_F \sigma_0)$ is the radius of unit optical depth
and $\dot N_{\rm ion}$ is the ionizing photon flux.  
Equation (9) then simplifies in
\begin{equation} 
\dot N_{\rm line}(t) =
{Y\over 2} \dot N_{\rm ion} \int_{-1}^{+1} d\mu \,\, 
\int_{R_{\rm min}}^{R_{\rm max}} {dr\over r_F} \, {\rm e}^{-r/r_F}
\end{equation} 
Finally, the solution of the above double integral can be expressed 
in terms of exponential integral function of second order, $E_2(x)$, as 
\begin{equation} 
\dot N_{\rm line}(t) \, =\,  
Y\, \dot N_{\rm ion} [1-E_2(t/2t_F)]
\end{equation} 
for $t\le \Delta t$, and 
\begin{equation} 
\dot N_{\rm line}(t) \, = \, Y\, \dot N_{\rm ion} 
[E_2((t-\Delta t)/2t_F)-E_2(t/2t_F)] 
\end{equation} 
for $t\ge \Delta t$.  
The time $t_F$ is defined as $t_F=r_F/c$.  
Note that
$\int_0^\infty dt \dot N_{\rm line}=\Delta t \dot N_{\rm ion}$.

The resulting light curve shows a power law growth for a time 
$t \le \Delta t$, when the line flux reaches its maximum, 
then slowly decays 
for a time scale of order $t_F$.  
After that time, the line flux decays exponentially.  
Very roughly, one can imagine the light curve of the line to be steady
at level $Y \dot N_{\rm ion} \Delta t/t_F$ for a time $t_F$, then going rapidly
to zero.  
If the ionized sphere were density bounded (i.e., the optical depth
$\tau_0$ of the scattering cloud is $<1$), rather than radiation bounded as
assumed so far, the fluorescence line flux would be at the same level as seen
above, but the duration of the line emission would be shortened by a factor
$\tau_0$.  
In Fig. 5 we show some illustrative examples of the iron line light curve.  
Note the slow power law decay of the line flux ($\propto t^{-1/10}$).

\subsection{Detectability of the iron fluorescence line}

The detection with instruments such as BeppoSAX, ASCA, AXAF or JET-X crucially
depends on the X--ray background collected within the detection cell.  The
BeppoSAX and ASCA Point Spread Function (PSF) is too large (with an Half Energy
Width, HEW, of the order of a few arcmin) and results in a strong background
which overwhelms the emission line.  
The good PSF of JET-X (HEW $\sim 25''$)
makes this instrument able to detect the line
and determine its redshift in
$\sim 100-200$ ks (for an assumed line flux of a 
few$\times 10^{-16}$ erg cm$^{-2}$ s$^{-1}$), even if 
at a low significance level.  
The limiting factor is the effective area at the line energy.  
For high redshift GRB afterglows, AXAF
BI CCDs provide a larger effective area (see above), so that the line can be
easily detected in about 100 ks (assuming the same flux).  
In the case of XMM a firm detection is obtained in the same amount of time 
and a determination of the line width is possible.

\section{Discussion}

We have shown in a simplified but analytical way that if gamma--ray bursts
originate in a dense environment their X--ray spectra are modified by
absorption, and the imprinted edges can be used to determine their redshifts.

Available information on the amount the absorption column in X--rays are scarce,
due to the fact that large field of view instruments detecting the gamma--ray
bursts (such as the WFC on BeppoSAX or BATSE on CGRO) have limited energy
resolution and/or they are insensitive to soft X--ray photons,
while the narrow field instruments can follow the X--ray afterglow
only some hours after the gamma--ray event, when the X--ray flux has presumably
faded away.  Despite these difficulties, for the very bright burst GRB980329 it
was possible to analyze the spectrum of the X--ray afterglow and the best fit
requires a column 
$N_{\rm H}=(1.0\pm 0.4)\times 10^{22}$ cm$^{-2}$ 
(with $z=0$, in't Zand et al 1998), much larger than the galactic value 
($N_{\rm H}^{gal}\sim 9\times 10^{20}$ cm$^{-2}$).  
Note that if the absorption is located at the
redshift of GRB980329, then the column is $(1+z)^3$ times larger 
(Paczy\'nski 1998).

Other indirect evidence of X--ray absorption comes from the extinction estimated
in the optical, from the fact that not all the bursts with a detected X--ray
afterglow could be detected in the optical, and to explain very hard X--ray to
gamma--ray spectra of the burst themselves.

Due to the fact that the oxygen and iron edges are the most prominent, there are
two almost distinct accessible part of the redshift--column density plane, 
one
characterized by a moderate $N_{\rm H}\sim 10^{21}$--$10^{22}$ cm$^{-2}$ and
$z\sim 0.1$--$0.5$ (assuming a detector energy window cutting-off at 0.1 keV),
and the other one characterized by a column larger than $10^{23}$.  Note that
for a standard dust to gas ratio, a column of $N_{\rm H} =10^{22}$ cm$^{-2}$
corresponds to an optical extinction $A_V\sim 6$ mag, precluding the possibility
to detect the optical afterglow.  
If, furthermore, it were not possible to perform spectroscopic optical 
observation of the host galaxy (either because a precise position is lacking, 
or because it is too faint), then the X--rays could be the only mean to 
determine the redshift, especially for strong bursts located at $z>2$--3, 
for which the iron edge lies in the most sensitive energy 
range of X--ray detectors.

For a detailed spectral analysis to be at all possible, we need large count
statistics, implying that the X--ray afterglow (assuming its temporal decay
follows the observed law also soon after the burst event) 
must be followed as soon as possible (i.e. within minutes).

A different case is instead possible if a sizeable fraction of the burst high
energy emission is re-emitted in the form of the fluorescent 
$K_\alpha$ iron line at an
intrinsic energy between 6.4 and 6.9 keV.  
In this case, possible if the
absorbing material has $N_{\rm H}>10^{23}$ cm$^{-2}$, one can detect the
fluorescence iron line up to $\sim 1-2$ years after the burst event.  
The key factors are the telescope effective area, to increase 
the throughput, as well as the HEW, to limit the background.  
The predicted flux for strong bursts is at
the level of a few 10$^{-16}$ erg cm$^{-2}$ s$^{-1}$, 
concentrated in the energy range [6.9--6.4]$/(1+z)$, 
which is detectable with the planned missions such as
Spectrum X--Gamma, AXAF, XMM and of course Constellation--X.

The case of GRB980425 (Soffitta et al. 1998), possibly associated with the
supernova 1998bw (Wang \& Wheeler 1998), deserves a specific discussion, due to
its proximity.  
We cannot exclude the possibility that the gamma--ray burst
emission is, in general, anisotropic, and that GRB980425 is a rare (because
fainter) event of a burst seen misaligned (Ghisellini et al.  1998).  
Causes of anisotropy could be different amount of baryon loading at different 
angles, together with (but not necessarily) different energy deposition as a 
function of angle (e.g.  Rees 1998).  
GRB980425 may then correspond to a burst seen with high baryon loading, 
small bulk Lorentz factor, lower total energy.  
In this
case the Doppler time contraction is much less extreme than in normal
bursts, resulting in a slow evolution of the afterglow light curves, while the
duration of the gamma--ray event, being controlled by the primary engine, is
unchanged.  This may explain the normal duration of the $\gamma$--ray emission,
the fact the optical flux reached its maximum long after the  burst event,
with a rising light curve, and the overall observed energetics.  
This idea
necessarily implies a much larger energy (of the order of $10^{51}$ erg, or
larger) being emitted at a different angle.  
If the environment of GRB980425 is
dense enough, then it is possible to detect a very strong fluorescence iron
line, with a flux of the order of $10^{-13}$ erg cm$^{-2}$ s$^{-1}$, easily
detectable by already flying satellites.  
Due to the importance of this
detection, we plan to propose X--ray observation in the direction of the
supernova 1998bw, even if there are not clear signs of absorption in the optical
spectrum.

\newpage

\setcounter{figure}{0}
\begin{figure} 
\psfig{file=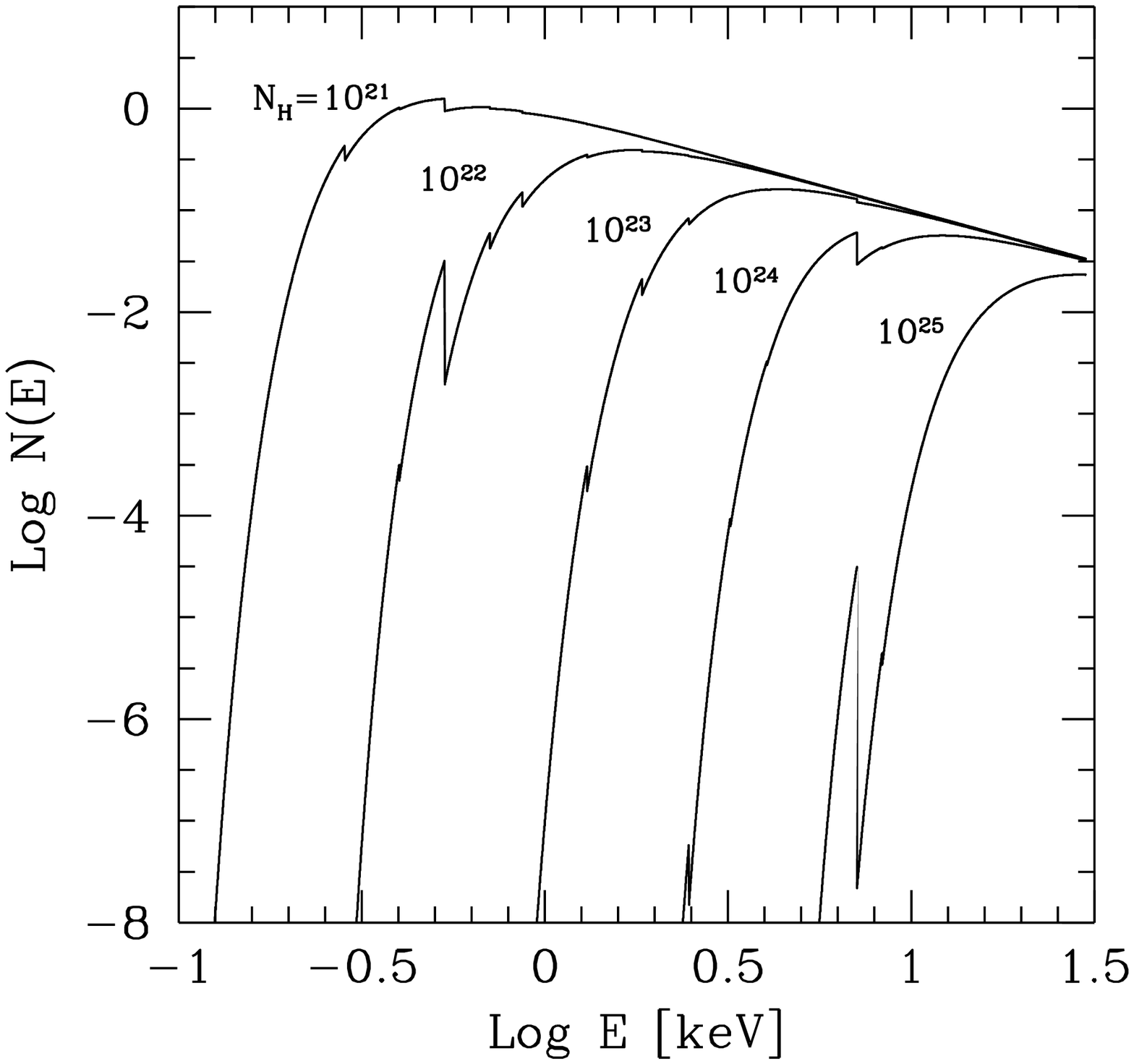,width=15 truecm,height=15truecm}
\caption[h]{The effect of different amount of absorption on a power 
law spectrum with photon index $\Gamma=1$.  
Labels indicate the values of $N_{\rm H}$ used.
For this figure, we have assumed solar abundances of metals and
that the absorption takes place at $z=0$.}
\end{figure}
\begin{figure} 

\psfig{file=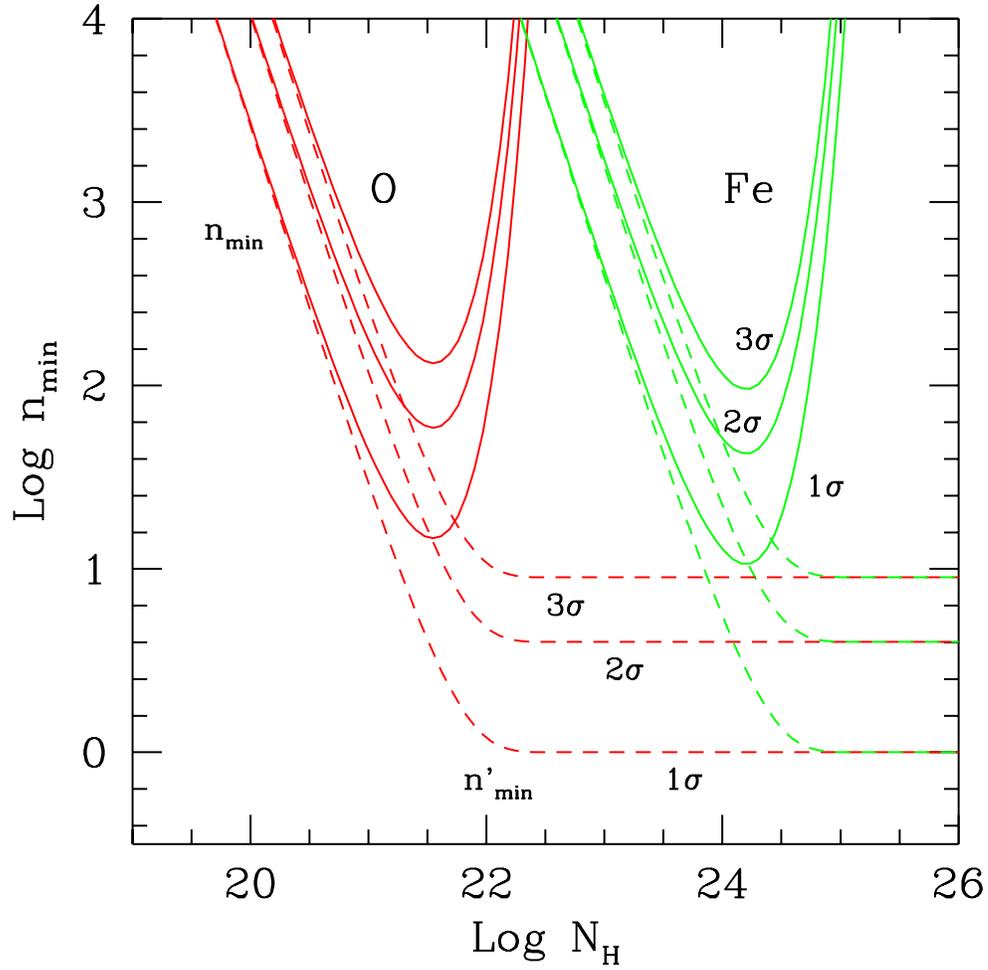,width=15 truecm,height=15truecm}
\caption[h]{The counts required to detect (at 1, 2 and 3 sigma) the edge of
neutral oxygen and neutral iron, as a function of the column density 
$N_{\rm H}$.  The solid lines refer to the unabsorbed counts, while the 
dashed lines refer to the counts actually measured.
Solar abundances of metals have been assumed.}  
\end{figure}

\begin{figure} 
\psfig{file=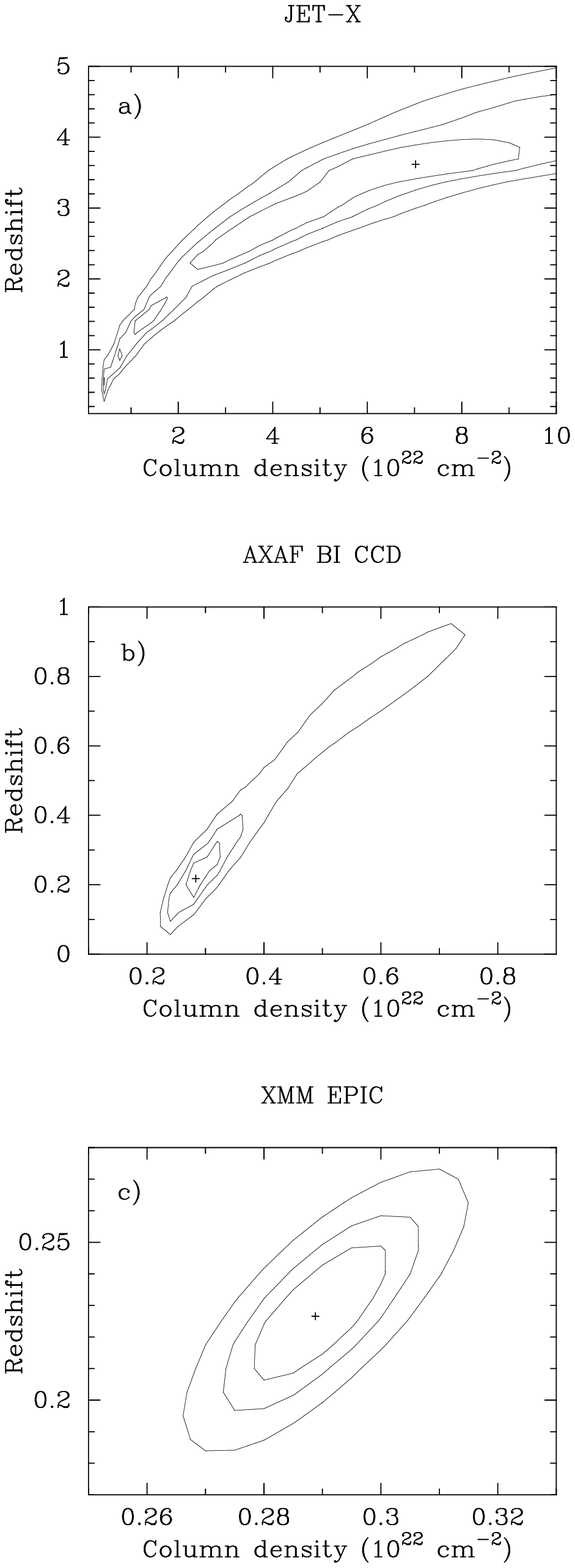,height=17 truecm} 
\caption[h]{Column density -- redshift contour plots for different 
X--ray instruments.  The input model has
$z=0.25$ and $N_{\rm H}=3\times 10^{21}$ cm$^{-2}$.  
The upper panel (a)  shows the
case of JET-X (two telescopes).  The middle panel (b) presents the case
of AXAF with BI CCDs and the lower panel (c) the case of XMM (three
telescopes).  See text for more detail on the input parameters of the
simulations.}
\end{figure}

\begin{figure} 
\psfig{file=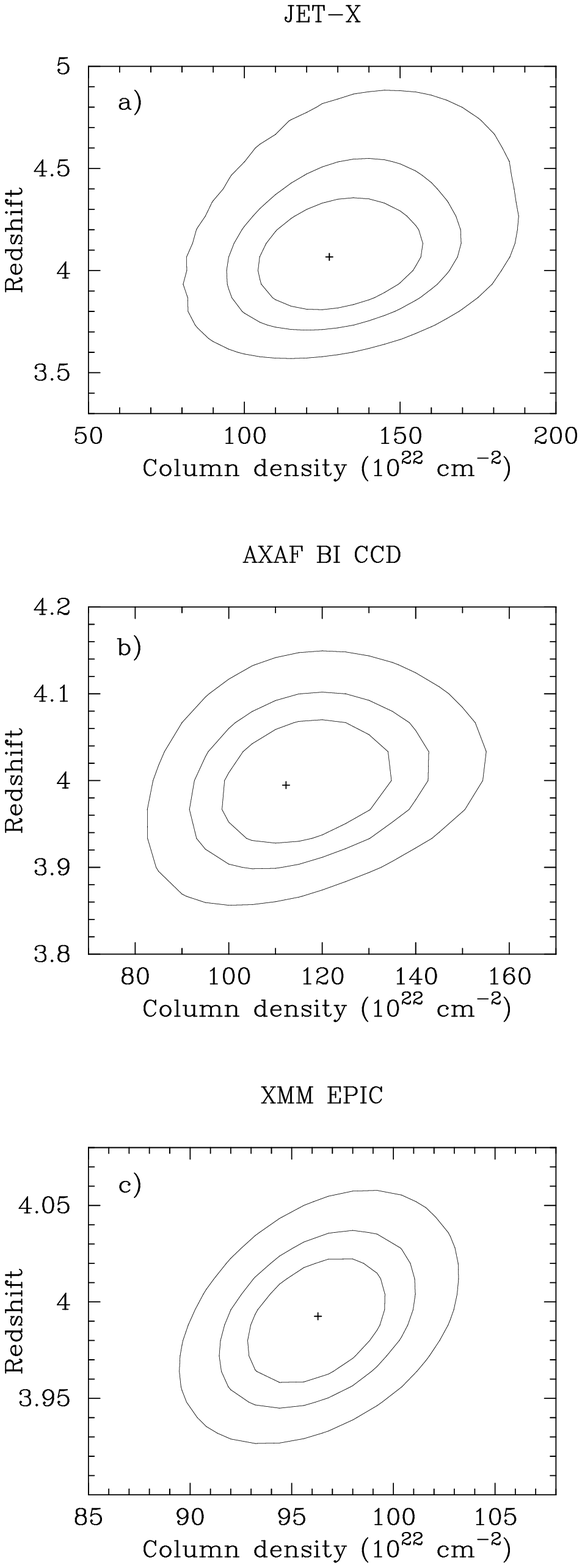,height=17 truecm} 
\caption[h]{Column density -- redshift contour plots for different 
X--ray instruments.  The input model has
$z=4$ and $N_{\rm H}=10^{24}$ cm$^{-2}$.  
The upper panel (a)  shows the
case of JET-X (two telescopes).  The middle panel (b) presents the case
of AXAF with BI CCDs and the lower panel (c) the case of XMM (three
telescopes).  See text for more detail on the input parameters of the
simulations.}
\end{figure}

\begin{figure} 
\psfig{file=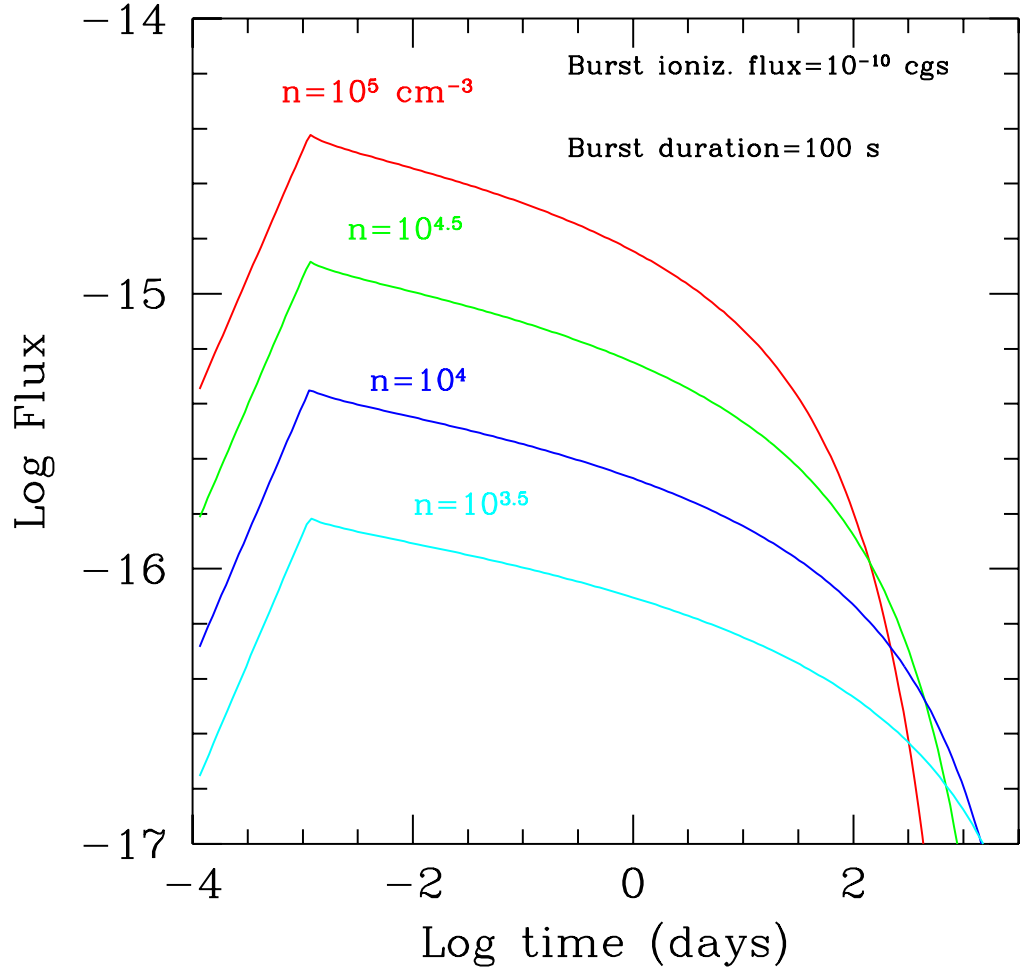,width=15 truecm} 
\caption[h]{The iron fluorescence line light curve
predicted for different values of the number density
of hydrogen, assuming a solar abundance of iron and a burst
of an ionizing flux of $10^{-10}$ erg cm$^{-2}$ s$^{-1}$
lasting for 100 seconds.}
\end{figure}
\end{document}